# Effect of Weak Impurities on Conductivity of Uniaxially Strained Graphene

Igor Yu. Sagalianov, Yuriy I. Prylutskyy
Taras Shevchenko National University of Kyiv
Dept. of General Physics, Dept. of Biophysics
UA-03022 Kyiv, Ukraine
isagal@ukr.net, prylut@ukr.net

Taras M. Radchenko, Valentyn A. Tatarenko
G.V. Kurdyumov Institute for Metal Physics of N.A.S.U.
Dept. of Metallic State Theory
UA-03142 Kyiv, Ukraine
tarad@imp.kiev.ua, tatar@imp.kiev.ua

*Abstract* — The main goal of our study was investigation of the influence of the deformations (sufficiently large for the establishing the non-zero gap) on electrotransport properties of impure graphene. To achieve this purpose, we implemented the simulation package that allows to perform numerical calculation of the conductivity and mobility of graphene samples subjected to two types of the uniaxial strain: along zigzag and armchair edges. All numerical calculations are performed within the Kubo–Greenwood methodology along with a tight-binding model. Transport properties are studied in case of a presence of impurity atoms described by the weak short-range scattering potential. Various deformation values are considered. The uniaxial strain acts as an additional source of the electron scattering, herewith, can strongly affect both mobility and conductivity of graphene and introduce anisotropy of its electron transport properties.

## I. INTRODUCTION

Graphene sheets, as well as other $2D$ flakes like elemental monolayers (silicone, phosphorene, etc) or various metal-dichalcogenides ($MoS_2$, $MoSe_2$, $WS_2$ and others), are widely considered as materials of the new generation for electronic devices [1–8]. Their extraordinary physical properties mutually arise from $2D$ structure and slowly degrade with enlarging of the number of layers in the flake. Each of monolayer materials has its own extraordinary features: pristine graphene has huge charge carrier's mobility and conductivity, phosphorene has initial band gap and sustains extreme (up to 35%) deformations, metal-dichalcogenides flakes consist from three layers (metal layer is sandwiched between two chalcogenide layers) hence exhibit their special quasi-$2D$ features etc. On the other hand, each of these materials demonstrates individual disadvantages caused by the same source, which causes their extraordinarily, — $2D$ structure and special chemical intra-layer bounding.

In case of graphene, its biggest and the most important lack is a band gap absence. Several alternative ways have been proposed for solving this problem [4–7]. The most common of them are as follows: formation of the nanoribbons, perpendicular electromagnetic field applied to the bilayer graphene, ordered spatial distributions of (substitutional or adsorbed) impurity atoms or various types of the strain.

$Sp_3$ hybridization and therefore strong chemical bounding between atoms in the graphene lattice form its extraordinary mechanical properties. Young's module exceeds $\approx 1$ TPa [5, 6] and tends to enhancement with additional defects [5, 6]. All that leads to the big value of the critical deformation. According to theoretical predictions, graphene can be successfully stretched up to $\approx 27.5$–$30\%$ without breaking of the lattice structure. Tension of graphene causes significant modification of its electronic structure. Deformation of graphene can be used as an effective method for the band-gap engineering and can be classified into several types, e.g., uniaxial, shear, and local strains. Depending on direction, there are zigzag and armchair uniaxial strains. It was shown that zigzag uniaxial deformation leads to the appearance of the band gap. However, the minimal required value needed for the gap to be opened is $\approx 23\%$, which lies near the critical limit of graphene's tensile strength. The larger deformation results to the wider gap. Such dependence has a linear character. One of the possible ways for decreasing critical deformation required for the band-gap appearance is introduction of the spatially-ordered point defects (impurities), which contribute to the band gap. Even minor concentration of defects allows us to decrease the minimal tensile strain required for the band-gap opening to $\approx 12.5\%$ or even smaller [9].

On the other hand, the exact influence of the impurities on the electron transport properties of graphene under strain is still almost undiscovered and requires additional studies. Even minor concentrations of the randomly distributed impurities can transform band gap into the pseudo-gap with nonzero concentration of the charge carriers inside. Moreover, impurities act as the scattering centres, hence, can directly affect electron transport properties of graphene.

## II. NUMERICAL PROCEDURE

We model graphene lattice within the framework of tight binding approximation. Interactions between different atoms in the Hamiltonian $\hat{H}$ are restricted to the nearest-neighbours,

$$\hat{H} = -u\sum_{i,i'=1}^{N} c_i^\dagger c_{i'} + \sum_{i=1}^{N} V_i c_i^\dagger c_i , \qquad (1)$$

where, $c_i^\dagger$, $c_i$ are the standard creation, annihilation operators acting on a quasiparticle at the lattice site $i$, $u = 2.7$ eV [10, 11] is hoping integral for the neatest-neighbour atoms occupying $i$ and $i'$ sites, and $V_i$ is on-site potential which is used to model scattering of the weak impurities. Including interactions in other coordination spheres (for example next-to-next nearest-neighbour approximation) causes minor influence on results of the simulations but drastically increases computational efforts. Since main goal of the present study is investigation of the huge graphene samples (which consist of several millions of sites) with small (realistic) concentration of defects, we conclude that nearest-neighbour approximation is appropriate for numerical simulations and acts as a golden middle between costs and effectiveness [12–16].

We introduce point defects (impurity atoms) into the honeycomb lattice via the scattering potential

$$V_i = \sum_{j=1}^{N_{imp}} V_j \delta_{ij}, \quad (2)$$

where $N_{imp}$ is the number of impurities on graphene sheet. It is important to notice that this type of scattering potential describes neutral impurities and can be used for analysis of the transport properties in graphene with short-range disorder. Thereby, during all our calculations, on-site potential was set to $\approx u$ (hopping parameter of the graphene lattice). Utilizing of the other types of disorder can significantly influence the results of the simulations. An absence of the periodic boundary conditions can moderately decrease conductivity in case of a finite computational domain. For the smaller lattice to be used, the larger boundary scattering are observed that strongly affects results.

After the tridiagonalization procedure, Hamiltonian matrix is reduced to the tridiagonalized form by passing to a new basis. Initial population of the lattice sites occurs with random-phase wave function:

$$|\psi_{rand}\rangle = \frac{1}{\sqrt{\Delta N}} \sum_{i=1}^{\Delta N} e^{2i\pi\alpha_i} |i\rangle \quad (3)$$

(where random phase $\alpha_i \in [0,1]$ and $|i\rangle = c_i^\dagger |0\rangle$) and transform Hamiltonian via passing on to a new orthogonal basis $\{r_i\}$, where set constructed wave function as the first element in the new basis, $|1\rangle = |\psi_{rand}\rangle$.

Density of electronic states at each energy level $E$ reads as

$$\rho(E) = \mathrm{Tr}\left[\delta(E - \hat{H})\right]. \quad (4)$$

In order to get total density of states we have to summarize all local densities of states for all sites,

$$\rho(E) = \sum_{i=1}^{N} \rho_i(E), \quad (5)$$

where $N$ denotes number of sites in the lattice, and $\rho_i$ can be calculated as imaginary part of the Green's function as follows below

$$\rho_i(E) = -\frac{1}{\pi} \mathrm{Im}\, G_{ii}\left[E + i\chi\right], \quad (6)$$

where $\chi$ is a smoothing coefficient. The smaller value of $\chi$ are used during the calculations, the more accurate however time-consuming results are obtained.

Time-dependent diffusion coefficient $D(E,t)$ relates with the position operator in Heisenberg representation, $\hat{X}(t)$, and time ($t$) evolution operator, $\hat{U}(t)$:

$$D(E,t) = \frac{\Delta\langle \hat{X}^2(E,t)\rangle}{t}, \quad (7)$$

$$\langle \Delta \hat{X}^2(E,t)\rangle = \frac{\mathrm{Tr}[(\hat{X}(t) - \hat{X}(0))^2 \delta(E - \hat{H})]}{\mathrm{Tr}[\delta(E - \hat{H})]}, \quad (8)$$

$$\hat{X}(t) = \hat{U}(t)^\dagger \hat{X} \hat{U}(t), \quad (9)$$

$$\hat{U}(t) = \exp\left(-\frac{i2\pi H t}{h}\right). \quad (10)$$

Finally, with utilizing of the diffusive transport regime (saturation of the $D(E,t)$ with time), we can calculate semi-classical conductivity $\sigma$ as follows below:

$$D_{max}(E) = \lim_{t \to \infty} D(E,t), \quad (11)$$

$$\sigma = \frac{e^2 \rho(E) D_{max}(E)}{\Omega}, \quad (12)$$

where $\Omega$ is area of the graphene simulation sample. All conductivity and mobility calculations have been provided alongside zigzag-edge direction in the graphene lattice.

Common technique in modelling of the uniaxial strain is introduction of the exponential decay for all hopping integrals in the Hamiltonian. We have used decay rate obtained from the independent experiment observations $\beta \approx 3.37$ in order to bound relations between old and new bond lengths ($l/a$) and hopping integrals ($u_{new}/u_{old}$) [5, 6]:

$$u_{new} = u_{old} e^{-\beta\left(\frac{l}{a} - 1\right)}. \quad (13)$$

Obviously, including of the Poison's ratio is necessary in order to get valid and qualified results. We have used Poisson's ratio $\nu \approx 0.15$, which is selected between the values measured for graphite [17] and calculated for graphene [18].

The most time consuming procedure is performing of the tridiagonalization procedure for calculation of the electronic density of states $\rho$ (DOS) and evaluation of the time evolution of the initial wave packet initially positioned on graphene surface. There are several methods designed for the enhancing of the calculation speed, like truncation of the continued fraction, careful detecting of the convergence procedure during time evolution and others. In addition, application of the fast programming languages combined with optimized mathematical libraries is necessary in order to get sufficiently good computational efforts. Thereby, we have developed effective programming package on Python combined with C-written mathematical libraries.

All calculations (density of electronic states, electron conductivity and mobility) were carried out using computer resources of the computational cluster located in the Taras Shevchenko National University of Kyiv.

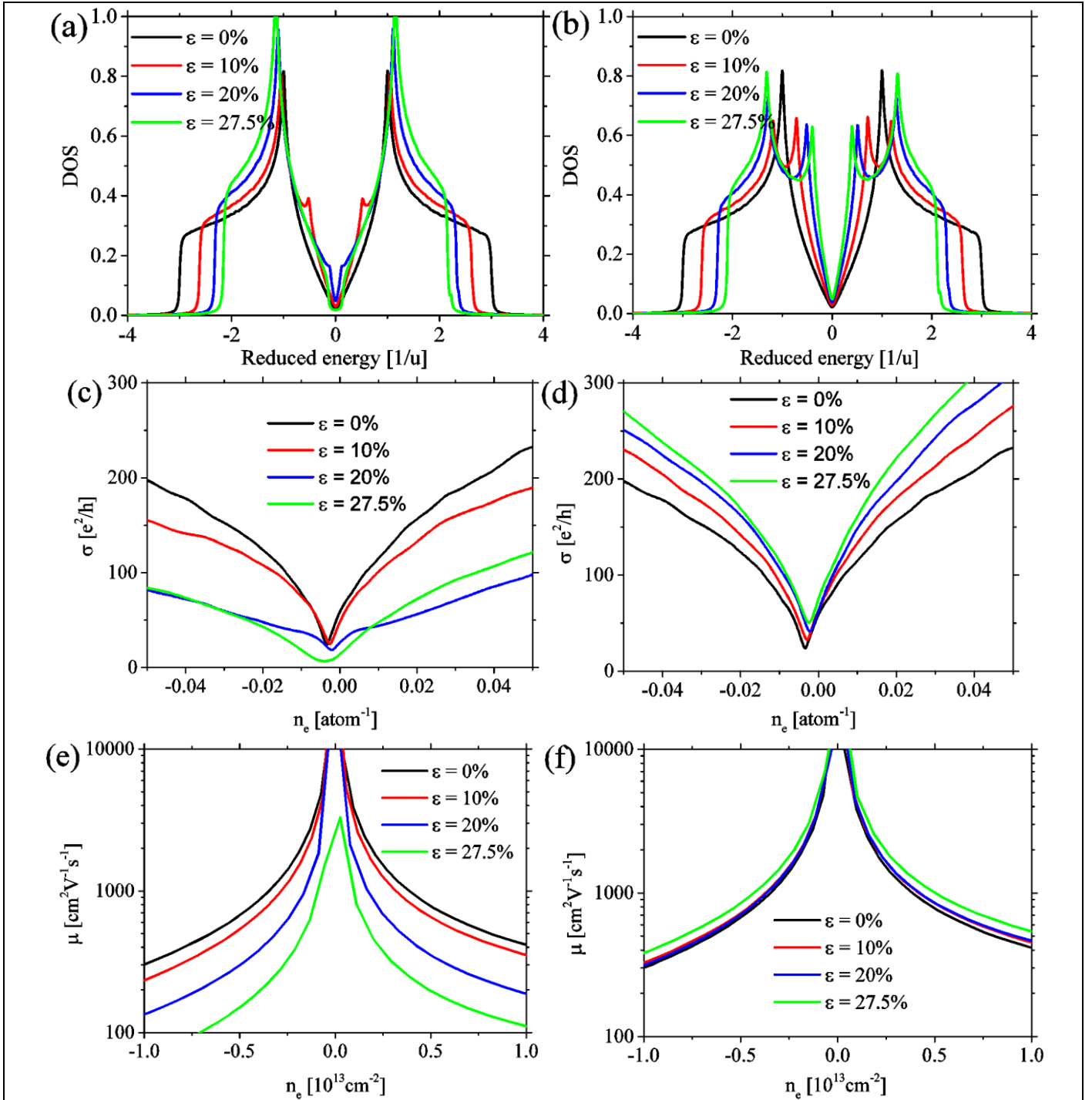

Fig. 1. Electronic and transport properties of impure graphene under the uniaxial strain, where deformation values vary from 0% up to 27.5% and impurity concentration $c = 0.1\%$. (a), (b) Densities of electronic states (in units of hopping integral $u$) for impure graphene under zigzag (a) and armchair (b) strains. (c), (d) Deformation-dependent conductivity for zigzag (c) or armchair (d) uniaxially-strained graphene. (e), (f) Mobility of the charge carriers in impure graphene under zigzag (e) or armchair (f) uniaxial strain. All conductivity and mobility calculations are carried out along zigzag-edge direction.

## III. CALCULATED RESULTS

All numerical simulations are performed on computational domain composed of graphene sheet, which consists of 1.7 millions of sites (atoms): 1700 and 1000 sites along zigzag and armchair directions, respectively.

Calculated density of electronic states for pristine (defect-free) graphene (not shown here) coincides with well-known curves in the literature [2, 4, 10, 11]. However, zigzag (Fig. 1a) and armchair (Fig. 1b) strains cause significant redistribution of the electronic states, especially in the vicinity of the neutrality (Dirac) point. Extreme zigzag deformation $\varepsilon = 27.5\%$ (Fig. 1a) causes appearance of the band gap, but the presence of 0.1% of weak impurities transforms it into the "quasi- (or pseudo-) gap" — plateau-shaped deep minimum.

To analyze the electron transport properties of graphene layer, we considered unchanged concentration of the impurities (0.1%) along with scattering on the boundaries of the sample in order to simulate realistic graphene sample. Since conductivity and mobility were calculated along the zigzag direction ($x$), we can conclude that stretching of the graphene along the $x$-direction significantly degrades its electrotransport properties (Fig. 1c). Conductivity decreases more than 25% for $\varepsilon = 10\%$ and more than on 50% for $\varepsilon = 20\%$. In addition, despite the presence of the polar impurities, electron–hole asymmetry does not appear significantly with stretching of the graphene even at the extreme deformations $\varepsilon = 27.5\%$. On the other hand, stretching in perpendicular ($y$) direction slightly compress graphene in the $x$-direction due to the Poison's ratio, hence we can see small enhancement of the conductivity in zigzag direction with strain in armchair one (Fig. 1d).

Next observation concerns mobility of the charge carriers. Stretching of the graphene in the $x$ direction causes anticipated decrease of the mobility both for electrons and holes more than 10 times (Fig. 1e). Contrary to this, stretching of the graphene lattice along the perpendicular direction almost does not affect mobility, even at extreme deformations of 27.5%. One of the possible explanations of this is redistribution of the charge carriers in Fig. 1b. Hence, various combinations of the direction of the uniaxial strains and direction of the propagation of charge carriers can be treated as an effective source of tuning the electron transport in graphene.

## IV. CONCLUSIONS

Uniaxial strain can serve as an effective source of tuning electrotransport properties of graphene layers. Studies of both electronic and transport properties require more careful consideration of the directions along the uniaxial strain is applied and the electron wave packet propagates.

In case of the unstrained graphene, there is no difference in the conductivities along $x$ and $y$ directions ($\sigma_{xx}$ and $\sigma_{yy}$) due to the huge size of realistic graphene lattices. However, in case of graphene subjected to the uniaxial strain, the latter introduces electron transport anisotropy and provides possibility of both almost full elimination or conserving of the mobility in graphene simultaneously with redistribution of the electron states in its energy spectra.